# Anomalous magneto-elastic and charge doping effects in thallium-doped BaFe$_2$As$_2$

Athena S. Sefat,[1,*] Li Li,[1] Huibo B. Cao,[2] Michael A. McGuire,[1] Brian Sales,[1] Radu Custelcean,[3] David S. Parker[1]

[1] *Materials Science & Technology Division, Oak Ridge National Laboratory, Oak Ridge, TN 37831*
[2] *Quantum Condensed Matter Division, Oak Ridge National Laboratory, Oak Ridge, TN 37831*
[3] *Chemical Sciences Division, Oak Ridge National Laboratory, Oak Ridge, TN 37831*

[*] Email: *sefata@ornl.gov*

Within the BaFe$_2$As$_2$ crystal lattice, we partially substitute thallium for barium and report the effects of interlayer coupling in Ba$_{1-x}$Tl$_x$Fe$_2$As$_2$ crystals. We demonstrate the unusual effects of magneto-elastic coupling and charge doping in this iron-arsenide material, whereby Néel temperature rises with small x, and then falls with additional x. Specifically, we find that Néel and structural transitions in BaFe$_2$As$_2$ ($T_N = T_s$ = 133 K) increase for x=0.05 ($T_N$ = 138 K, $T_s$ = 140 K) from magnetization, heat capacity, resistivity, and neutron diffraction measurements. Evidence from single crystal X-ray diffraction and first principles calculations attributes the stronger magnetism in x=0.05 to magneto-elastic coupling related to the shorter intraplanar Fe-Fe bond distance. With further thallium substitution, the transition temperatures decrease for x = 0.09 ($T_N$ = $T_s$ = 131 K), and this is due to charge doping. We illustrate that small changes related to 3$d$ transition-metal state can have profound effects on magnetism.

## Introduction

While the reasons for high-temperature superconductivity (HTS) remain unsolved, its manifestation in a particular antiferromagnetic material ('parent') [1] is an equal mystery. Following the discovery of HTS in spin-density-wave (SDW) BaFe$_2$As$_2$ ('122') by K-doping [2-4], experimental work for finding superconductivity in similar structural materials was pursued [5]. In iron-based superconductors (see refs. [6-18] for several reviews), there is a highly complex interplay between a number of factors such as the apparent competition between magnetism and superconductivity that occurs upon charge doping of parent antiferromagnetic compounds, the close proximity of the tetragonal-to-orthorhombic lattice distortion to the antiferromagnetic temperature and the associated nematicity, the competing hypotheses of local moment vs. itinerant magnetism, and numerous other factors including orbital ordering. Within a year after the initial discovery of superconductivity in iron-arsenide materials, it was realized that magneto-elastic coupling, or the coupling of magnetic properties to strain, was very strong [19-22]. Indeed this is clear from the very existence of the stripe antiferromagnetic ground state, in which (nearest neighbor) Fe atoms 2.78 Å apart in the Fe plane order ferromagnetically while next nearest neighbor Fe atoms order antiferromagnetically, despite being only additionally 0.02 Å apart.

In this work we demonstrate and discuss, based on experimental work and first principles calculations, a situation in which magneto-elastic coupling is sufficiently strong to *reverse* the well-documented reduction of Néel temperature with chemical substitutions. We find that with Tl doping of BaFe$_2$As$_2$ (Ba$_{1-x}$Tl$_x$Fe$_2$As$_2$; Tl-122) with x=0.05, the Néel temperature increases from $T_N$ = 132 K to 138 K, and that further doping of x=0.09 reduces $T_N$ to 131 K. This is contradictory to the conventional wisdom regarding the iron arsenides, in which the $T_N$ decreases with doping of the parent. In order to understand this behavior, we compare our Tl-122 system to the known behavior of the isoelectronic hole-doped system Ba$_{1-x}$K$_x$Fe$_2$As$_2$ (K-122) [23,24], in which a uniform reduction in $T_N$ with K chemical-substitution is observed. Both K and Tl are hole-dopants, based on Hall effect measurements. We should note that the radii of 8-fold coordinated Tl$^+$ (159 *pm*) is similar to Ba$^{2+}$ (142 *pm*) [25]. Both K (151 *pm*) and Tl should be monovalent dopants and essentially isoelectronic in $BaFe_2As_2$. As is well known for K doping, the antiferromagnetism is diminished and a superconducting region starts for $x \sim 0.1$, and the maximum $T_c$ is 38 K for $x \sim 0.4$ [26,27]. For us, Tl doping of more than $x \sim 0.1$ in $BaFe_2As_2$ crystal was not possible, and hence potential superconductivity is not seen. In this study, we focus on the disparate behavior of the antiferromagnetic ordering temperature in both Tl-122 and K-122 cases, and explain it as a 'strain' effect, originating in the differing response of the $BaFe_2As_2$ lattice to the particular dopant atom. For understanding magnetism in iron-arsenides, and as early as 2008 [28], it was noted that local density approximation (LDA) gives better agreement with experiment, compared with the generalized gradient approximation (GGA) that overstates the magnetic order [29]. Indeed, the recent work [30] for different 122 materials using reduced Stoner theory show better agreement with experimental magnetism. This is a report of the highly unusual magnetic behavior in $Ba_{1-x}Tl_xFe_2As_2$ in that $T_N$ increases with Tl (x=0.05), and then decreases (x = 0.09), and report of structure, thermodynamic and transport experimental data. Theoretical LDA calculations show a strong link of magnetism to the fine details of crystal structure.

**Results & Discussions**

$Ba_{1-x}Tl_xFe_2As_2$ crystals had sheet/block morphologies and dimensions of $\sim 5 \times 4 \times 0.1$ mm$^3$ or smaller in $a$, $b$, and $c$ crystallographic directions, respectively. Similar to $BaFe_2As_2$, the crystals formed with the [001] direction perpendicular to the plates. See **Fig. 1**. For finding the level of x in Tl-122, more than 10 spots (~90 μm diameter each) were averaged on 3 random crystal pieces in each as-grown batch; the standard deviation was derived from the measured data variation. The measured EDS values are $x_{EDS}$=0.045(10) for $x_{nominal}$= 0.15, and $x_{EDS}$=0.091(21) for $x_{nominal}$= 0.30. EDS analyses indicated that less Tl was substituted in the crystal than the amount added in solution, in accordance to all literature chemical-doping synthesis results [31]. In this manuscript, the crystals are denoted by x= 0, 0.05, and 0.09 to describe all thermodynamic and transport bulk properties.

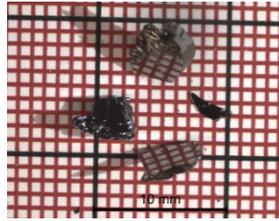

**Figure 1:** FeAs-grown single crystals of $Ba_{1-x}Tl_xFe_2As_2$.

The list of refined structural data from single-crystal X-ray diffraction is listed in **Table 1**. For $BaFe_2As_2$, the interatomic Fe-Fe distance $d_{Fe-Fe}$ = 2.8006(7) Å ($=a/\sqrt{2}$) and As internal coordinate (0,0,0.3537) are comparable to literature values [2]. Here are a few observations from this Table. Although $c$-lattice parameter expands in Tl-122 probably due to larger Tl$^+$ pushing apart FeAs layers, the overall unit cell volume shrinks. For $x_{nominal}$=0.15, the expansion of the $c$-parameter is +0.9%. The increase in interlayer distances between FeAs layers is evident from interplane As-As bond expansion. Tl in 122 promotes more direct Fe-Fe intraplanar interactions, evident from smaller $a$-axis and shorter $d_{Fe-Fe}$. From the relative changes in both tetrahedron angles of FeAs$_4$, it seems that Tl-doping promotes a more of regular tetrahedral coordination (109.5°) around Fe. Arsenic height, $z_{As}$, expands slightly. Shorter intraplanar Fe-Fe bond distance is expected to cause higher $T_N$ through magneto-elastic coupling, especially since states near the Fermi level are derived almost exclusively from Fe orbitals [32]. By Tl substitution, although there is little to no change in the $z$ position of arsenic, there is slight decrease of the Ba/Tl-As distance. This trend along with the decreases of $a$, increase of $c$, and decreases of unit cell volume are comparable to the literature report of Tl-doping of $BaCu_2Se_2$ with the same crystal structure [33]. X-ray refinement of Tl atomic occupancy on the one single crystal piece chosen out of $x_{nominal}$=0.15 batch is $x_{X-ray}$ =0.070(13), and that on an $x_{nominal}$ =0.30 crystal piece is $x_{X-ray}$=0.07(2). There may be an evidence of slight decrease of $c$-lattice parameter with more Tl-doping, possibly an indication of mix of smaller Tl$^{3+}$. Overall and within error, the refined Tl-doped single-crystal X-ray diffraction structures are identical for the two crystal pieces, each selected at random from each batch. According to our EDS analyses of many crystals above, such refined Tl-occupancies fall within the expected $x_{EDS}$=0.045(10) and $x_{EDS}$=0.091(21), respectively.

Single-crystal X-ray diffraction data are collected on an ~100 μm size crystal; hence, we believe that EDS x averaging (described above) gives the best representative 'x' relevant to describing bulk physical behavior.

Table 1: Structural refinement parameters for BaFe$_2$As$_2$ and two randomly chosen pieces from each of Ba$_{1-x}$Tl$_x$Fe$_2$As$_2$ with x$_{nominal}$=0.15 and 0.30 batches. Single-crystal X-ray diffraction data were collected and refined [33] at the temperature above the magnetic/structural transitions at 173(2) K.

|  | x=0 | x$_{nominal}$=0.15 | x$_{nominal}$=0.30 |
|---|---|---|---|
| $a$ (Å) | 3.961(1) | 3.938(1) | 3.939(2) |
| $c$ (Å) | 12.968(3) | 13.089(3) | 13.086(7) |
| $V$ (Å$^3$) | 203.42(4) | 203.0(1) | 203.1(2) |
| Fe-Fe (Å) | 2.8006(7) | 2.7847(7) | 2.785(1) |
| As-Fe-As (°) | 108.40(3), 111.63(6) | 108.81(3), 110.80(6) | 108.83(5), 110.8(1) |
| Fe-As (Å) | 2.394(1) | 2.3922(9) | 2.393(2) |
| As-As (Å) | 3.794(3) | 3.827(3) | 3.824(5) |
| arsenic $z$ coordinate; As height (Å) | 0.3537(1); 1.345 | 0.3539(2); 1.360 | 0.3538(1); 1.358 |
| Ba/Tl-As (Å) | 3.382(1) | 3.378(2) | 3.379(2) |
| Tl occupancy | 0 | 0.070(13) | 0.07(2) |

The magnetic susceptibility ($\chi$) behaviors of Ba$_{1-x}$Tl$_x$Fe$_2$As$_2$ crystals are anisotropic (**Fig. 2a**), as expected from the tetragonal structure. In fact, $\chi_{ab}/\chi_c$(300 K) = 1.3 for x= 0 with $c/a$ = 3.27, and increases to 1.5 for x = 0.09 that has $c/a$ = 3.32. The $\chi$ increases with temperature with no evidence of a maximum, or Curie-Weiss behavior; Tl-122 behavior is very similar to that observed in SDW CrAs [35] and parents of $A$Fe$_2$As$_2$ with $A$=Ca, Sr, and Ba [34], which are multiband semi-metallic materials consisting of both electron and hole Fermi surfaces that nest to give SDW upon cooling. From $d\chi/dT$ analyses, x=0 gives a peak at 132 K, x=0.05 gives 138 K and 140 K anomalies, and x=0.09 gives a peak at 129 K. The low temperature susceptibility results increase below these anomalies (below ~ 100 K), probably indicating magnetic correlations due to Tl substitution. This low-temperature upturn is featured more dramatically in 'polycrystalline' samples of BaFe$_2$As$_2$ and K-doped BaFe$_2$As$_2$, and the causes are not clear. From magnetic susceptibility results, we found our first evidence of *an increase of $T_N$ in* BaFe$_2$As$_2$ ($T_N$ ≈ 132 K) with chemical substitution (x = 0.05) to $T_N$ ≈ 138-140 K, before decreasing at $T_N$ ≈ 129 K for x= 0.09.

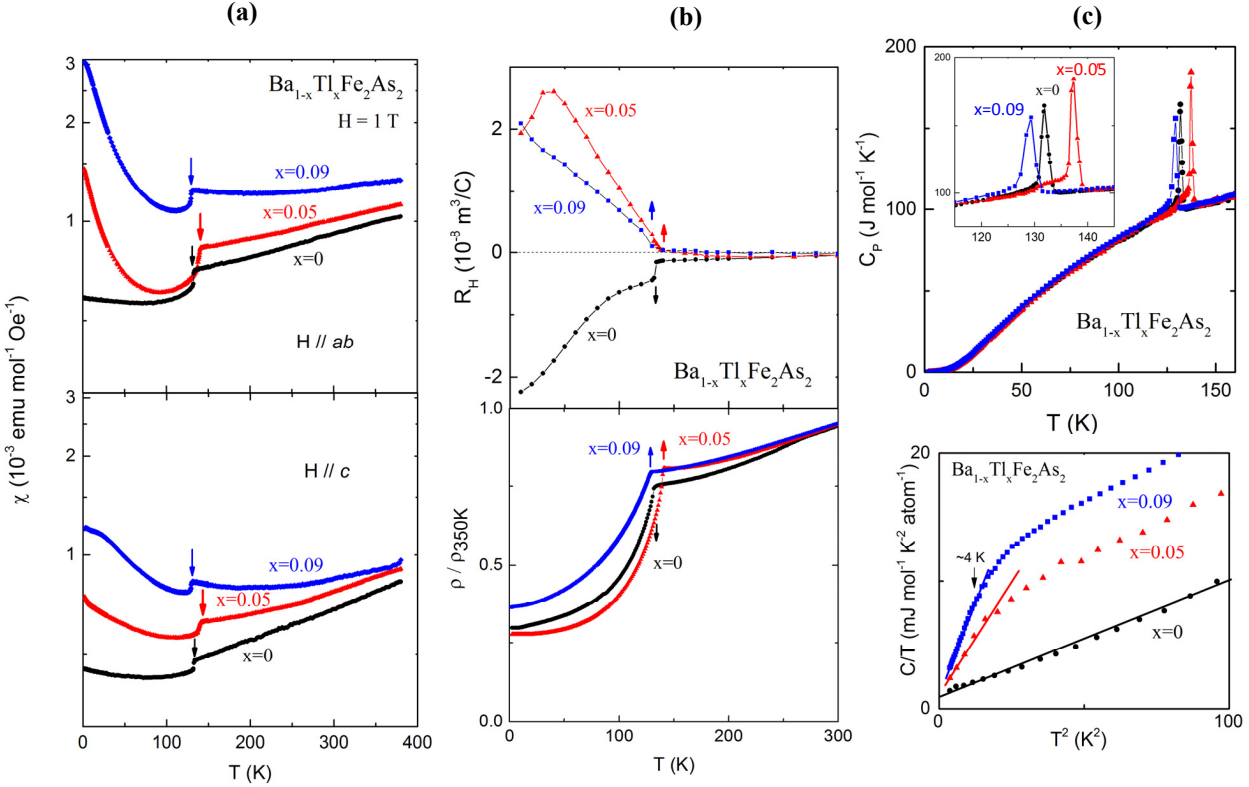

**Figure 2:** For $Ba_{1-x}Tl_xFe_2As_2$ crystals with x = 0, 0.05, and 0.09, temperature dependence of (a) magnetic susceptibility along different crystallographic axes, (b) Hall effect and electrical resistivity, and (c) specific heat. The plot of $C/T$ versus $T^2$ is shown below 10 K in (c) bottom.

We should note that there are varying experimental reports on the values of $T_N$ in $BaFe_2As_2$ parent, presumably caused by off-stoichiometry, unintentional doping, chemical order, variable levels of strain, or even different inferred $T_N$ methods. For the parent, we find $T_N$= 132(1) K (FeAs flux-grown crystals) [here, and for example 32, 36-39], although there are reports of $T_N$ = 137-139 K [40, 41], $T_N$= 136(1) K (Bridgman grown) [42, 43], $T_N$= 140 K (polycrystalline samples) [2], $T_N$= 85 K (Sn-grown crystals, with small Sn substitution) [44], and even superconductivity at ~ 22 K (Sn-grown crystals) [45]. In one of our recent manuscripts, we found that taking a piece of as-grown crystal ($T_N$= 132 K) and thermally-annealing it (at 700 °C for 30 days) produces $d\chi/dT$ anomaly at $T_N$ = 137(1) K [35]. Although the specific physical reasons for such an increase in $T_N$ with prolonged thermal-annealing were not found, it was speculated as structural strain relief. The fact that $T_N$ increases for Tl-122 at 5% value, is opposite to such an expectation as we are increasing disorder/inhomogeneity with chemical substitution. In this present manuscript, we emphasize that we only study as-grown crystals, made using the same synthesis method.

The resistivity of Tl-122 is not high, due to high densities of states at the Fermi level. The resistivity of Tl-doped samples is slightly less than the parent ($\rho_{300\ K}$< 1 mΩ.cm), probably due to increase in carriers by hole doping. All samples display similar metallic behavior with ρ decreasing upon cooling (**Fig. 2b** bottom). The resistivity falls rapidly, probably at the SDW transition for each sample, with an expected periodic modulation in the density of the electronic spin with $2\pi/q$ spatial frequency, similar to $BaFe_2As_2$ [46]. Each resistive transition temperature can be estimated from the peak in $d\rho/dT$; the inferred values are 133 K for x=0, 140 K for x =0.05, and 128 K for x=0.09. The Hall voltage was calculated from the antisymmetric part of the transverse voltage (perpendicular to the applied current) under magnetic-field reversal at fixed temperature. Shown in **Fig. 2b** top, is the temperature dependence of the Hall coefficient, $R_H$. Hall data is characteristic of a metallic system with both electron and hole bands at the Fermi level.

Although the carrier amount $n$ may be inferred via $n = 1/(qR_H)$, interpretation of $R_H$ is complicated by the multiband nature. The temperature-dependence of $n$ may be related to changes in relaxation rates associated with different electron and hole pockets, and its drop below $T_N$ is consistent with Fermi-surface gapping. For $BaFe_2As_2$, $R_H$ values are negative between 300 K and 10 K; it drops abruptly below 133 K with an increasing magnitude with decreasing temperature, due to more contribution from higher mobility electron carriers. For Tl-122, $R_H(T)$ behavior are roughly the same, indicating similar physics. The values are negative between 300 K and 200 K, then change sign and become positive close to and below structural transition; for x=0.05, there is an abrupt rise below 140 K. This indicates that for Tl-122, the accompanied transition to SDW phase increases the hole-like Fermi surface volume. For x=0.09, the hole carrier contributions may be less than x=0.05 below $T_N$, perhaps due to slight $Tl^{3+}$ mixing.

For $BaFe_2As_2$, heat capacity increases below 134 K peaking at 132 K (**Fig. 2c**), comparable to single crystal data from literature [e.g. 39] associated with $T_s$ and $T_N$. For x= 0.05 Tl doping, C(T) develops a peak below 140 K peaking at 137 K. For x= 0.09, it rises below 132 K with the highest point at 129 K. The transition width of heat capacity peak for x= 0.09 is twice as that for x=0.05 and it may be due to chemical non-uniformity for higher Tl-doped levels. For x=0.09, the peak starts to form below 131.4 K. The plot of C/T versus $T^2$ is displayed up to 10 K and is linear for $BaFe_2As_2$, with Sommerfeld-coefficient of $\gamma = 5.9(3)$ mJ $K^{-2}$ $mol^{-1}$ and Debye temperature of $\theta_D = 297$ K, fitted below 6 K, comparable to literature [37]. Although Tl-doped samples have less of a linear region (fits below 4 K), γ values are comparable to $BaFe_2As_2$, probably because the DOS for these materials is governed by iron arsenide layers, not spacer chemical details. The excess heat capacity in Tl-doped samples may be due to magnetic disorder and related to magnetic susceptibility upturns at low temperatures.

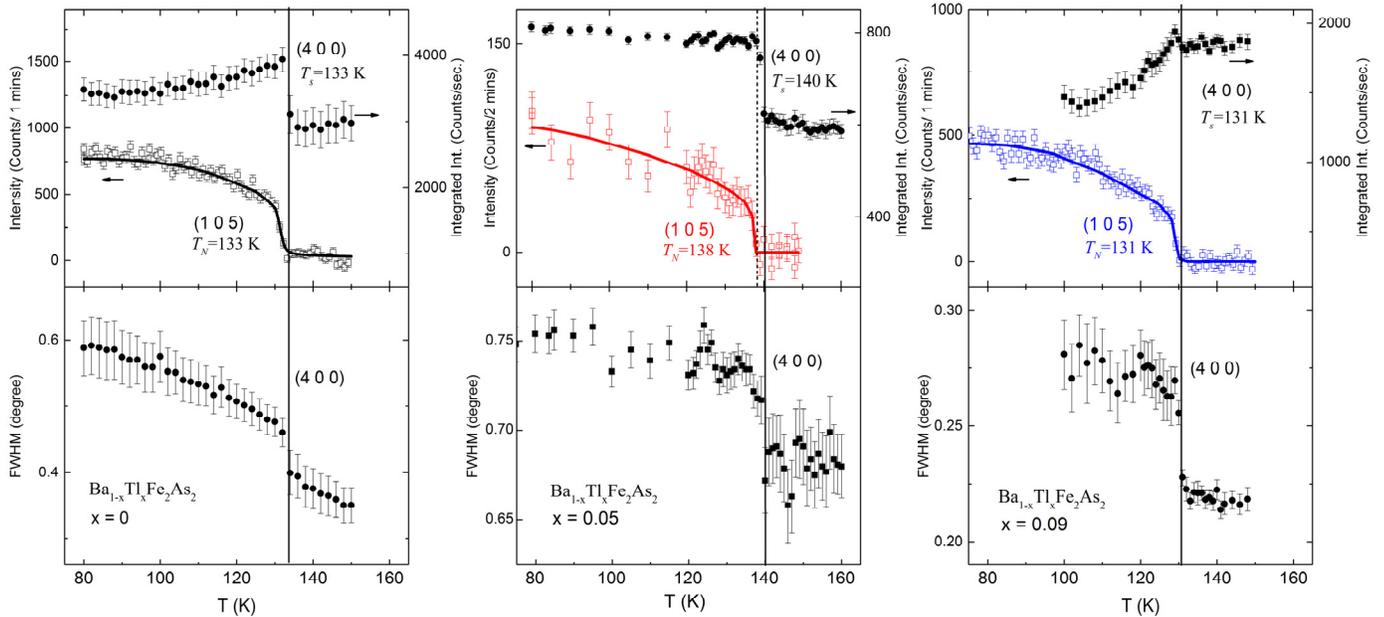

**Figure 3:** For $Ba_{1-x}Tl_xFe_2As_2$ crystals, neutron diffraction results for x=0 (left), x=0.05 (middle), and x=0.09 (right). The temperature dependence of Bragg reflections are shown upon warming. Top panels: integrated intensity of the nuclear peak $(400)_O/(220)_T$ and the peak intensity of the magnetic peak $(105)_O/(½½5)_T$; curved line is a guide for eyes. The solid/dashed line marks the structural/magnetic transition. Bottom panels: full peak width of $(400)_O/(220)_T$ at half peak maximum.

According to neutron diffraction data (**Fig. 3**), for the parent there is a simultaneous structural transition [47] and a magnetic transition to a SDW phase. In this magnetic state, the spins are aligned along *a*-axis; the nearest-neighbor (nn) spins are aniparallel along *a*- and *c*-, and parallel along shortest *b*-axis. The nesting ordering wavevector is $q = (101)_o$ or $(½ ½ 1)_T$, relative to the tetragonal (T) or orthorhombic (o) nuclear cells [47]. For $BaFe_2As_2$, the reported ordered moment is ~0.9 $\mu_B$/Fe [47, 48]. We expect that the ordered-moment for Tl-122 will not deviate significantly from such reports, as thallium is substituted for Ba between iron arsenide layers. Similar to $BaFe_2As_2$ parent, the nuclear peak of $(220)_T$ is most likely to split to $(400)_O$ and $(040)_O$ orthorhombic Bragg reflections below $T_s$ in these lightly Tl-doped $BaFe_2As_2$ crystals. **Fig. 3** and top panels show the relative integrated intensity of structural $(400)_O/(220)_T$ and magnetic $(105)_O/(½½5)_T$. The increased intensity of the structural peak is due to reduced extinction effect by the structural transition from tetragonal to orthorhombic space group. The peak intensity decrease after the intensity jump (release of the extinction effect) at $T_s$ for the parent compound and x=0.09 doped one are due to further peak broadening. The temperature-dependence of full peak width at half maximum (FWHM) of $(400)_O/(220)_T$ is shown in the bottom panels; the peak broadening indicates that the tetragonal (220) peak becomes separated as (400) and (040) peaks. Here we find evidence that for x=0.05, $T_s$ and $T_N$ are split at 140 K and 138 K, respectively. However, $T_N = T_s =$131 K for x=0.09, slightly lower than $T_N = T_s = $ 133 K for the parent.

We performed theoretical calculations in order to understand the highly unusual behavior of doping with thallium for which $T_N$ first increases in 5% chemical substitution, then decreases with 9% in $BaFe_2As_2$. This behavior is opposite to that expected and found for $Ba_{1-x}K_xFe_2As_2$ [23,24]. Theoretically, the role of doping on the magnetism in $BaFe_2As_2$ using K was previously analyzed [49]; this publication found an increase in the magnetic ordering energy, defined as the difference in total energy between the stripe phase and the checkerboard state. Such an increase in ordering energy would, in the simplest mean-field approximation, have association with an increase in $T_N$ that is not observed in K-122. Note that both in the previous work and in the current work the stripe and checkerboard phases are found to have lower total energy than the non-magnetic case, so that in a mean-field approximation $T_N$ is governed by the difference in energy between these two phases. One possible reason for the previously reported increase in ordering energy with K-doping was the use of the GGA [28], which can overstate the tendency towards magnetic order. Indeed more recent work [30] used "reduced Stoner theory", for which the spin-dependent part of the exchange correlation potential is reduced by a constant factor, and showed better agreement of calculated magnetic properties with experiment. An alternative approach, which we adopt here, is to use the LDA and compare the differences between hole-doped systems of K- and Tl-122. Coupled with this choice is the specific choice of the crystal structure: we use the low-temperature experimental orthorhombic structure (lattice parameters and arsenic height), as we refined by powder X-ray diffraction data on $BaFe_2As_2$ and Tl-122, and those listed in Ref. [24] for K-122; see **Table 2**. For the base $BaFe_2As_2$, the structure is nearly identical to that optimized via the GGA in the previous work [49].

**Table 2:** The experimental structural details used for LDA calculations for Tl-122, and that of K-122 reported in ref. [24]; refined in *Fmmm* space group.

|  | T(K) | *a* (Å) | *b* (Å) | *a/b*-1 | *c* (Å) | arsenic z coordinate |
|---|---|---|---|---|---|---|
| $BaFe_2As_2$ | ~ 2 K | 5.6157(2) | 5.5718(2) | 7.87 x $10^{-3}$ | 12.9424(4) | 0.35375(3) |
| $Ba_{1-x}Tl_xFe_2As_2$ (x=0.09) | 15 K | 5.5973(2) | 5.5607(2) | 6.57 x $10^{-3}$ | 13.0024(4) | 0.35414(7) |
| $Ba_{1-x}K_xFe_2As_2$ (x=0.1) | ~ 2 K | 5.5997(1) | 5.5587(1) | 7.38 x $10^{-3}$ | 13.0031(4) | 0.35405(3) |

**Table 3**: Theoretically LDA calculated energy difference ΔE, in meV per Fe atom, between the ground-state striped and the checkerboard structures.

|  | ΔE, meV/Fe | Percent change in |ΔE| from $BaFe_2As_2$ |
| --- | --- | --- |
| $BaFe_2As_2$ | -47.5 | 0 |
| $Ba_{1-x}K_xFe_2As_2$ (x=0.07) | -43.8 | -7.8 |
| $Ba_{1-x}K_xFe_2As_2$ (x=0.1) | -42.4 | -10.7 |
| $Ba_{1-x}Tl_xFe_2As_2$ (x=0.07) | -47.5 | 0 |
| $Ba_{1-x}Tl_xFe_2As_2$ (x=0.1) | -46.1 | -3.0 |

We model three physical structures – the base $BaFe_2As_2$ compound, and those of the Tl and K substituted compounds, for mid and higher substitution levels. In order to isolate the role of charge doping, we used the same structure of the respective ~10% dopants (**Table 2**). **Table 3** depicts a great difference in magnetic ordering energy behavior between the K-122 and Tl-122. In the K-doped case ΔE drops sharply, reducing over 10 percent from the baseline value for x=0.1, while for Tl there is no reduction for x=0.07 and a slight reduction for x=0.10. Note that we have chosen charge doping levels of 0.07 and 0.1 for both K-doping and Tl-doping in order to permit a direct comparison between the two scenarios, as the previous study's [24] K-doping levels were 0.07 and 0.1. In a mean-field scenario, $T_N$ is proportional to this magnetic ordering energy, so that the disparate behavior of ΔE under K and Tl doping conditions should be mirrored by corresponding changes in the $T_N$. We depict this in **Fig. 4**, which shows the ΔE values in **Table 3**, normalized to the $BaFe_2As_2$ value, along with our measurements of $T_N$ of the Tl-122, again relative to that of pure 122. The theoretical results for the two dopants are clearly very different; while the theoretical results for Tl-122 do not reproduce the increase in $T_N$ for x=0.05, they do exhibit the slight decrease thereafter. We have checked that in $BaFe_2As_2$ the non-magnetic state lies higher in energy than both magnetic states so that these are not metastable states. The question then becomes: what is the reason for the great divergence in behavior of Tl doping and K doping? We note that while monovalent Tl in eight-fold coordination has a significantly larger radius than K, this is not reflected in the *c* lattice parameters, which from **Table 3** are virtually identical for Tl and K doping. Of more interest are *a* and *b* lattice parameters, which both decrease from the base value, but in different ways – for Tl-122, *a* is smaller than that for K-122 while for *b* the reverse prevails. Fe-Fe distances in the Tl structure are 2.7804 and 2.7986 Å, while in the K structure they are 2.7794 and 2.7999 Å, which exhibit no obvious trend. All in all, the results of the calculations suggest an extreme sensitivity of the magnetic results to small structural changes, which can be labeled as magneto-elastic effects known to be strong in the iron-arsenides. Indeed this should be immediately clear from the change in sign (based on the stripe ground-state) of the effective Fe-Fe magnetic interaction from ferromagnetic to antiferromagnetic with a change in Fe-Fe distance from 2.78 to 2.80 Å. Against this backdrop, it is more reasonable that changes in the Fe-Fe distances of order 0.001 Å can significantly affect magnetic properties. Another way to see this is to note that from **Table 2** the deviation of the planar lattice parameter ratio from unity, *a/b*-1, is some 12% larger for the K-122 than for Tl-122. Since magnetism in these compounds is closely tied to the structural transition, this is suggestive of significantly differing behavior in these two systems.

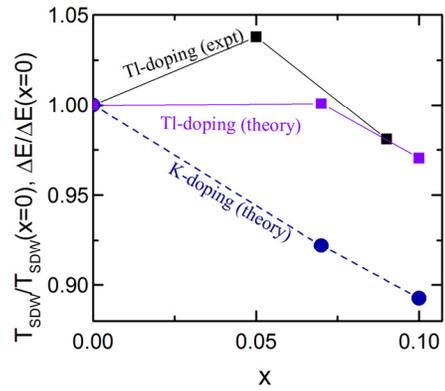

**Figure 4:** The normalized ordering energy calculated for Tl- and for K-122, and the normalized experimental $T_N$ for Tl-122 from data here.

In conclusion, through experimental measurements of electrical resistance, heat capacity, magnetic susceptibility, and neutron diffraction, we find that the anomalous behavior of the coupled $T_N = T_s$ in $Ba_{1-x}Tl_xFe_2As_2$ increasing and splitting for the x= 0.05 doped crystal at $T_N$ = 138 K and $T_s$ = 140 K. Further Tl-substituted x= 0.09 crystal gives reduced the value of $T_N$ = 131 K, with an overlap with $T_s$. Evidence from single crystal x-ray diffraction attributes the stronger magnetism for x= 0.05 to shorter intraplanar Fe-Fe bond distance. This is the first report of thallium-doping of $BaFe_2As_2$, and the first clear demonstration of the disparities between magneto-elastic coupling and charge-doping. Theoretical LDA calculations indicate the great sensitivity of the $T_N$ behavior to the crystal structural details.

**Methods**

*Synthesis and preparation.* Single crystals of $Ba_{1-x}Tl_xFe_2As_2$ were grown out of self-flux using the conventional high-temperature solution(flux)-growth technique [31]. The FeAs binary was synthesized similar to our previous reports [31,32]. Small Ba chunks, Tl chunks, and FeAs powder were mixed together according to the ratio Ba/Tl/FeAs= (1-x)/x/4 with $x_{nominal}$= 0, 0.15, and 0.30. Each mixture was placed in an alumina crucible. A second catch crucible containing quartz wool was placed on top of this growth crucible and both were sealed in a silica tube under ~1/3 atmosphere of argon gas. Each of these mixtures was heated for ~24 hours at 1180 °C, and then cooled at a rate of 2 °C/hour, followed by a decanting of the flux. For batches of $x_{nominal}$= 0 and 0.15, the decanting temperature at 1090 °C produced crystals. For $x_{nominal}$= 0.30, this temperature resulted in complete spin (no crystals), hence further cooling and centrifugation at 1050 °C was done.

*Elemental and structure analyses.* The chemical composition of the crystals was measured with a Hitachi S3400 scanning electron microscope operating at 20 kV, and use of energy-dispersive x-ray spectroscopy (EDS). Phase purity, crystallinity, and the atomic occupancy of crystals were checked by collecting data on a Bruker SMART APEX CCD-based single-crystal X-ray diffractometer, with fine-focus Mo $K_\alpha$ radiation. One piece of each crystal from $x_{nominal}$= 0, 0.15, and 0.30 batch (all sides less than 0.1 mm) was covered in Paratone N oil and kept under a stream of liquid nitrogen, at 173(2) K, above their potential structural or magnetic transition values. We used the initial atomic coordinates from the known $BaFe_2As_2$ structure, and then refined against the X-ray data using full matrix least-squares methods of SHELXTL software; absorption correction was applied using SADABS [33].

*Physical property measurements.* Magnetization of the samples was performed in Quantum Design (QD) Magnetic Property Measurement System with the field in the *ab*-plane and along the *c*-crystallographic axes. Each sample was cooled to 2 K in a zero-field, then the data were collected by warming to ~ 370 K in magnetic field of 1 Tesla. The electrical transport was performed in the Quantum Design Physical Property Measurement System (PPMS). Four electrical leads were attached to the samples using Dupont 4929 silver paste, with resistance measured in the *ab* plane. The temperature dependence of the heat capacity was also obtained using the QD PPMS, via relaxation method.

*Neutron diffraction.* Single crystal neutron diffraction was performed using the four-circle diffractometer HB-3A at the High Flux Isotope Reactor (HFIR) at the Oak Ridge National Laboratory, to distinguish between the structural and magnetic transitions in x=0, 0.05, and 0.09 Tl-122. The selected crystals were initially used for EDS and property measurements. The neutron wavelength of 1.542 Å was used from a bent perfect Si-220 monochromator [50].

*Theory.* Charge doping was simulated using the virtual crystal approximation (VCA), under the assumption that both the Tl and K substitute on the Ba-site and are monovalent. All calculations were performed using the all-electron planewave first principles density functional theory code WIEN2K [51] using the LDA. All calculations used a minimum of 1000 *k*-points in the full Brillouin zone and an $RK_{max}$ of 7.0, where $RK_{max}$ is the product of the minimum sphere radius and the maximum planewave vector. We have checked convergence of the energy differences with respect to both $RK_{max}$ (by increasing $RK_{max}$ to 9.0) and the number of *k*-points (by using approximately 5000 *k*-points); the differences in energy between the stripe and checkerboard structures changed by less than 0.1 percent in all cases. Our present calculations, similar to those reported, found an antiferromagnetic interaction along the *c*-axis for K-122, so that the ground-state magnetic ordering vector is (101).

**Acknowledgement**

This work was primarily supported by the U.S. Department of Energy (DOE), Office of Science, Basic Energy Sciences (BES), Materials Science and Engineering Division (A.S., M.M., B.S.), and Chemical Sciences, Geosciences, and Biosciences Division (R.C.). This study was partially funded (L.L., D.P., R.C.) by ORNL's Lab-Directed Research & Development (LDRD). The work at ORNL's High Flux Isotope Reactor (HFIR) was sponsored by the Scientific User Facilities Division, Office of BES, U.S. DOE (H.C.). We appreciate S. Kuhn's assistance in chemical composition (EDS) measurements.

**Author contributions**

LL prepared the samples, and carried out elemental and structure analyses, and physical property measurements. HC carried out neutron diffraction experiments. MM performed low-temperature X-ray powder diffraction. RC did single crystal X-ray diffraction. DP performed theoretical calculations. AS supervised the experiments, discussions, and wrote the paper. All authors reviewed the manuscript, and contributed to the analyses, and writing of the manuscript.